# A Novel Algorithm for Optimised Real Time Anomaly Detection in Timeseries

Krishnam Kapoor[1]

[1] Indian Institute of Technology, Kharagpur, India
`krishnamkapoor@iitkgp.ac.in`

**Abstract.** Observations in data which are significantly different from its neighbouring points but cannot be classified as noise are known as anomalies or outliers. These anomalies are a cause of concern and a timely warning about their presence could be valuable. In this paper, we have evaluated and compared the performance of popular algorithms from domains of Machine Learning and Statistics in detecting anomalies on both offline data as well as real time data. Our aim is to come up with an algorithm which can handle all types of seasonal and non-seasonal data effectively and is fast enough to be of practical utility in real time. It is not only important to detect anomalies at the global but also the ones which are anomalies owing to their local surroundings. Such outliers can be termed as contextual anomalies as they derive their context from the neighbouring observations. Also, we require a methodology to automatically determine the presence of seasonality in the given data. For detecting the seasonality, the proposed algorithm takes up a curve fitting approach rather than model based anomaly detection. The proposed model also introduces a unique filter which assess the relative significance of local outliers and removes the ones deemed as insignificant. Since, the proposed model fits polynomial in buckets of timeseries data, it does not suffer from problems such as heteroskedasticity and breakout as compared to its statistical alternatives such as ARIMA, SARIMA and Winter Holt. Experimental results the proposed algorithm performs better on both real time as well as artificial generated datasets

## 1 Introduction

Anomalies can be defined as observations that deviates so much from other observations as to arouse suspicion that it was caused by a different mechanism (Hawkins, 1980[1]). Anomaly detection in timeseries is an important area with respect to business. Timely alerts about anomalous behaviour of data can save millions of dollars worth of losses. Identification and subsequent treatment of anomalous data is also required for developing a reliable forecasting model. Although, an extensive amount of work has been done in this field[2], still none of the existing algorithms have the desired accuracy, performance or auto tuning capacity to be used in real time anomaly detection on business data. So, we present a novel algorithm which has a high degree of accuracy in indentifying local anomalies and is optimised enough to be of practical utility for real time data.

We adopted a local polynomial based approach which fits local regression models to small segments of out data. These local models are subsequently used to compute the residual time series which is then compared with an adaptive threshold to identify the anomalies. This adaptive thershold adjusts itself to detect anomalies which occur owing to the local context of the neighbourhood. We then employ a classifier in the form of a filter which determines if the detected anomaly is a false alarm. Any anomaly which does not satisfy the threshold criteria of this filter is discarded as a false positive. This enables the algorithm to effectively detect anomalies even in the presence of high noise.

Employing such a method to detect anomalies requires two very important input parameters – the size of local window in which the polynomial is fitted and the order of the polynomial to be fitted in this window. The first parameter can be determined by detecting the periodicity of the data. If the data under consideration turns out to be periodic, it is naturally a good idea to take the window size equal to the periodicity of the data. We empoly a non-parametric to detect the periodicity accurately. Our algorithm also handles problems such as loss of periodicity[*] and presence of multiple periodicities in the data. The second parameter can then subsequently be estimated from the calculated window size.

In this paper, we have evaluated the performance of 5 different outlier detection methods against our algorithm on various real as well as artificial data sets. The algorithms compared are based on different statistical and machine learning models such as SARIMA, Holt Winter, One Class Support Vector Machines and FB Prophet. Since, we did not come across any standard dataset tagged for outliers, we resort to comparison based analysis on differernt datasets which were manually annotated by industry experts.

---

[*] The data under consideration might become aperiodic for a short interval due to noise.

## 2 Related Work

We have divided various outlier detection algorithms present in the literature into four major categories-

### 2.1 Statistical Models

Business and economic time series are often complex and exihibit difficult periodic patterns. A large number of model based approaches (S. C. HILLMER and G. C. TIAO[3]) have been proposed which assumes time series to be a multiplicative or additive combination of components such as trend, seasonality and residue. Such temporal based decomposition methods go all the way back to Holt-Winters methodology (Chatfield 1978[4]). These approaches try to model each of these components seperately. However, these decompositions are often suited to the problem they are designed for and thus fail to generalize to all cases.

ARIMA based models (Chatfield 2016[5]) have also heavily been used to address the problem of outlier detection. In case of seasonal data, more generalized SARIMA models have been employed. The problem with these models is that these are linear combinations of Autoregressive and Moving Average terms whereas many business time series are non-linear in nature. Although, the problem of automatic determination of order has been addressed using autoarima[6], these models still suffer from problems such as heteroskedasticity and automatic periodicity determination. Apart from this, these statistical models assume that the timeseries is ergodic with a gaussian noise model which might not hold in every case.

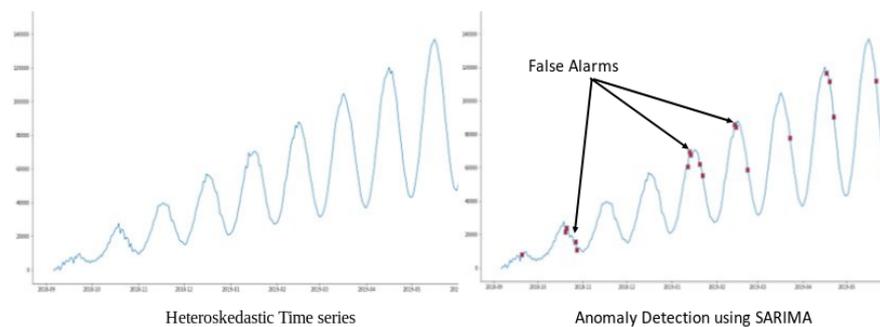

**Fig. 1.** Automated determination of periodicity (Source: Periodicity Paper)

### 2.2 Machine Learning Models

There is a vast literature machine learning based approaches to detect anomalies in time series data. For instance, the negative selection algorithm[7] is a interesting approach. Many neural network[8] based approaches have been tried and even combination of various machine learning and statistical models, for instance neural network and SARIMA[9] have been studied for the purpose of forecasting . This approach can as well be employed for the problem of outlier detection but the problem lies in the fact that time series data is usually short and hence neural networks are often not trained properly.

Anomaly classifiers based on One-class SVMs[10] are an attractive alternative because they can naturally detect outliers in a set of vectors. However, these models do not generalize to all scenarios and require the proportion of outliers in the dataset as a hyper-parameter, hence limiting their usage for real time data.

Netflix's anomaly detection based on Principle Component Analysis[11] and Twitter's S-H-ESD[12] are also promising alternatives. However, drawbacks such as requirement of high cardinality of data for PCA to work limit the usage of these algorithms for every time series.

### 2.3 Deep Learning Models

LSTM based models[8] can also be employed for anomaly detection but they too suffer from the same problems, such as lack of data for proper training and extremely slow operation, as neural networks. Hence, deep learning based models cannot be used in our case.

### 2.4 Curve Fitting Models

FB Pophet[13] proposes a non linear regression model with three model components: trend, seasonality and holiday. Prophet fits a linear or logistic function to model the trend and depends on Fourier series for seasonality detection. It also takes into account the effect of holidays to account for irregular shocks. Subsequently, Prophet uses LOESS regression for modelling the error terms. This model can be used for anomaly detection as well using the calculated bounds. Prophet was made keeping business time series into account, hence it was expected to perform well. However, our experiments show that it does not possess a reliable periodicity detection capacity for real time anomaly detection.

## 3 Outlier Detection Methods

### 3.1 Holt Winter And SARIMA

We have consider two statistical models for performance comparison with our algorithm, namely the Holt Winter Model[4] and SARIMA[9]. The Holt Winter model[4], also known as the triple smoothing approach, is a combination of Holt model which was formulated in 1957 and Winter's model in 1960. There are two versions of this model, the additive version and the multiplicative version. Both these versions are constitute three components – the level, the growth and the seasonal factor. We have used the additive version for this paper. On the other hand, SARIMA model is basically a seasonality generalization of the ARIMA model with 3 additional hyperparameters. It is denoted as SARIMA(p,d,q)(P,D,Q)$_m$ where p stands for trend autoregression order, d stands for trend difference order, q stands for trend moving average order, P stands for seasonal autoregressive order, D stands for seasonal difference order, Q stands for seasonal moving average order and m stands for the number of time steps for a single seasonal period.

In both the cases, we calculate the value of an observation using the specified model and subsequently calculate the residual errors by substracting this calculated value from the true value. If the deviation of any point in the residual errors is more than a set threshold[*], we mark that point as an anomaly. All the models considered in this paper are tested in both online[**] as well as offline[***] environments.

### 3.2  One Class Support Vector Machines

One Class Support Vector Machines[10], abbreviated as OC SVM, convert a time series into a set of vector in projected phase spaces and subsequently classify outliers based on these vectors. The proportion of outliers in the data given by hyperparameter ν[****] must be specified to the model. One of the obvious drawbacks of this model is that the value of this hyperparameter is not known in advance.

### 3.3  FB Prophet

FB Prophet fits a number of linear or a logistic trends to the given time series based on its detected changepoints. It is basically a non-linear additive model capable of handling yearly, weekly and daily periodicity along with integrating the effect of calendar holidays. One of its key features is that it is robust to missing data and possess automatic seasonality detection. The upper and lower confidence bands are calculated for the given data automatically and point violating these bands is marked as an anomaly. However, one of its major drawbacks is that it does not automatically detect any periodicity other than the predefined yearly, weekly and daily seasonalities.

## 4    The TOAD Model

The acronym TOAD stands for "Time Optimised Anomaly Detection". In our algorithm, we have broken down the problem of outlier detection into four sub-problems, namely periodicity detection, outlier identification, removal of false alarms and optimisation. Periodicity detection can be handled accurately by the use of a hybrid algorithm given by Vlachos, Michail & Yu, Philip & Castelli, Vittorio. (2005)[13]. According to this method, a time series has a true periodicity only if we get a peak in PSD and a local maxima in ACF corresponding to that periodicity.

---

[*] Here, we have taken the threshold to be 2 standard deviations from the mean of residual.
[**] Here, the algorithm is run multiple times with a new data point being introduced each time to check if the latest point is anomalous in real time.
[***] Here, the entire time series is fed to the algorithm at once to determine the anomalies.
[****] For this paper, we have used the value of ν which gives the best results.

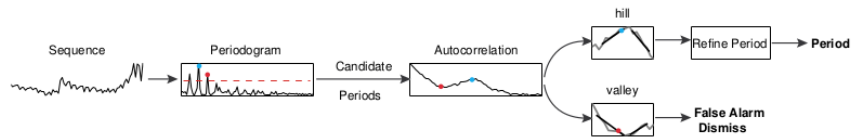

Fig. 2. Automated determination of periodicity (Source: [13])

We have observed that in some cases, periodicity detection becomes extremely hard for some part of the data due to the presence of extreme noise. In order to overcome this problem, we save the periodicity of the data once it is detected and adjust the value in case it changes. During our experimentation, we encountered another problem, in case of small time series data, we get hills towards the end of ACF. However, these candidates for periodicity so be removed by setting a confidence window to make sure that the detected periodicity is truly present. If multiple true periodicities are detected, we consider the smallest one. The reason for this will be clear once we discuss how curve fitted model works.

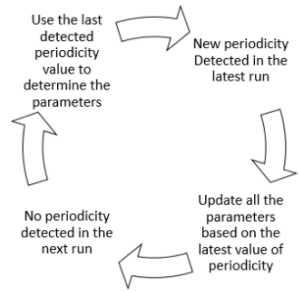

Fig. 3. Process to update the periodicity

### 4.1 Outlier Detection Methdology

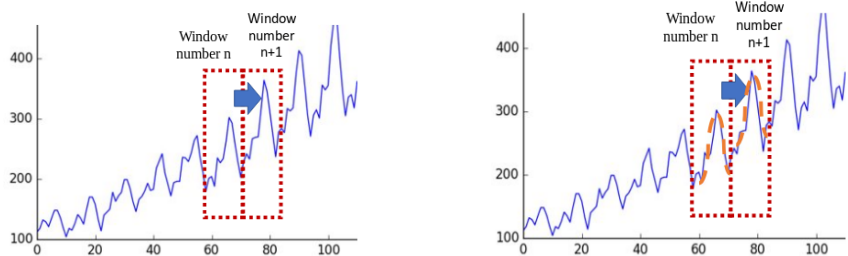

Fig. 4. The illustration shows the concept of bucketing and subsequently fitting local models to a time series

Once the periodicity is detected, we bucket the time series into a number of contiguous windows to fit local models. In case of periodic series, it is a good idea to take the periodicity of the data as the window size. This is also verified experimentally as shown in fig. 4. Our experiments show that increasing the polynomial order by 1 for every five data points results in a good fit of local models. In the case of non-periodic data, we experimentally conclude that a window size of 10 with a linear regression model will result in proper curve fitting local models. The process of bucketization is initiated from the first data point of the time series. If there are not sufficient data points in the last window, we merge that window with the previous one. All the local piece-wise models are then stacked together and subsequently the window breaks between them are smoothened by using polynomial regression with the same window size around the point of the window break. Hence, this results in a smooth trend replica of our time series data.

We obtain the residual errors by substracting this replica from the original time series. The residual errors are then bucketed in a similar fashion as the original data which subsequently help to calculate the local standard standard deviations. Bounds for each of the buckets are calculated by taking a linear combination of these local deviations with the global standard deviations given by the equation -

$$Residual\ Bounds = \alpha (SD_{Local}) + (1-\alpha)(SD_{Global})$$

where α is constant between 0 to 1. We set the threshold as two times the calculated bounds for a given bucket in both directions with respect to the mean of each bucket. The data points which reside outside these bounds are deemed as outliers.

### 4.2 False Alert Removal (FAS) Filter

We calculate the bounds for all the local models of the bucketed signal in the same way as we calculated bounds on the residual errors given as -

$$Signal\ Bounds = \alpha (Signal\ SD_{Local}) + (1-\alpha)(Signal\ SD_{Global})$$

We now define a filter known as FAS whose value is calculated for each for the these buckets by the formulae -

$$Filter = \log_{10}\left(\frac{Signal\ Bounds}{Residual\ Bounds}\right)$$

This value essentially measures the strength of the residual errors w.r.t. to the strength of the original signal. The idea behind designing such a filter is that many local anomalies become insignificant in the presence of a very strong trend or seasonality. This filter assess the relative strength of the residual errors w.r.t. the signal. We remove the outliers of a window if the FAS value of that window is greater than a threshold. A threshold of 1 implies that the variance of residual errors is 100 times weaker than the variance of the signal. Hence, a threshold value of 1 is reasonable for the FAS filter.

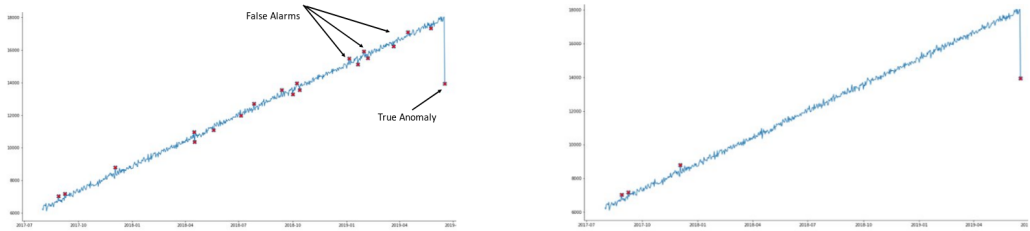

**Fig. 5.** The illustrastion on the left shows the results without employing the FAS whereas the image on the right shows the filtered output.

### 4.3 Hyperparameters

The values of α and threshold of the filter are the only hyperparameters which need to be given to the algorithm. During our experimentation, we found out that 1 is a suitable value for the filter for all practical purposes. Also, we propose that equal weightage be given to both the local and global standard deviations to avoid detectiong small spikes in very low variance regions. So, the value of alpha is taken to be 0.5.

### 4.4 Optimisation

The algorithm can optimised further for real time anomaly detection in a couple of ways. First, having a fixed window size of 10 in the case of non-periodic time series can result in a large number of buckets which increase the overall time complexity of the algorithm. Instead, it would be a better idea to fix the number of windows. Our experiments show that taking a window size equal to 10% of the length of the data and polynomial of degree 2 produces equally good results.

Another important question which arises here is that should the algorithm be run everytime a new data point is detected or is there method to reduce the number of runs. In order to tackle this problem, we propose a mechanism to select the ocassions on which the algorithm should be run. Since, the signal bounds of the last window from the previous run will be available, these bounds can be used to determine the run ocassions. The anomaly detection algorithm should be run only when the value of the time series violates these signal bounds in any of the directions. By employing this mechanism, we were able to speed up the algorithm upto 3 times as discussed in the next section.

## 5   Results

For analysing the relative performance of our algorithm, we compare it with 4 other anomaly detection methodologies. These include Holt Winter, SARIMA, One Class SVM and Prophet based anomaly detection algorithms. For getting a better understanding about performance of these algorithms, we resort to both online and offline based comparison. We start by comparing these algorithms on 10 artifically produced datasets of which 2 of the toughest cases are highlighted below. The first data set is made of an artificial sine wave with a periodicity of 28 and heteroskedastic amplitude as shown in fig 6. This dataset does not contain any anomaly. The second data set has 500 data points with an artificial breakout at the $250^{th}$ point as shown in fig 6.

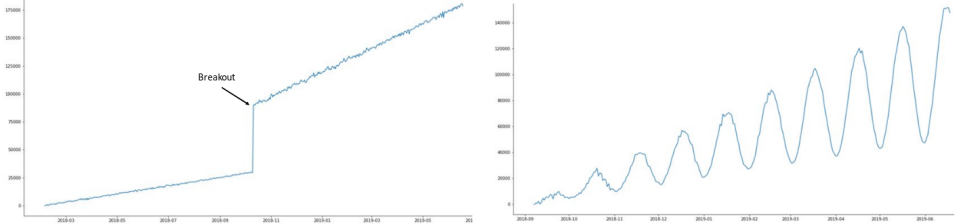

**Fig. 6.** The figure on the left shows the second dataset and the figure on the right shows the first dataset.

**Table 1.** Performance analysis of various algorithms in both online and offline environments on the first dataset.

| Algorithm | True Positives | True Negatives | False Positives | False Negatives | Type |
|---|---|---|---|---|---|
| TOAD | 0 | 256 | 0 | 0 | OFFLINE |
| Holt Winter | 0 | 239 | 17 | 0 | OFFLINE |
| SARIMA** | 0 | 237 | 10 | 0 | OFFLINE |
| OC SVM* | 0 | 254 | 2 | 0 | OFFLINE |
| Prophet | 0 | 249 | 7 | 0 | OFFLINE |
| TOAD | 1 | 255 | 0 | 0 | ONLINE |
| Holt Winter | 0 | 190 | 66 | 0 | ONLINE |
| SARIMA** | 1 | 229 | 26 | 0 | ONLINE |
| OC SVM* | 0 | 220 | 36 | 0 | ONLINE |
| Prophet | 0 | 249 | 7 | 0 | ONLINE |

---

* We have used value of ν=0.01 for OC SVM model.
** Here, we have employed SARIMA$(1,0,0)(1,0,1)_{28}$ model.

**Table 2.** Performance analysis of various algorithms in both online and offline environments on the second dataset.

| Algorithm | True Positives | True Negatives | False Positives | False Negatives | Type |
|---|---|---|---|---|---|
| TOAD | 1 | 499 | 0 | 0 | OFFLINE |
| Holt Winter | 0 | 474 | 25 | 1 | OFFLINE |
| SARIMA* | 0 | 494 | 5 | 1 | OFFLINE |
| OC SVM** | 0 | 493 | 6 | 1 | OFFLINE |
| Prophet | 1 | 483 | 16 | 0 | OFFLINE |
| TOAD | 1 | 494 | 5 | 0 | ONLINE |
| Holt Winter | 0 | 474 | 25 | 1 | ONLINE |
| SARIMA* | 0 | 493 | 6 | 1 | ONLINE |
| OC SVM** | 0 | 298 | 201 | 1 | ONLINE |
| Prophet | 1 | 473 | 16 | 0 | ONLINE |

Table 1 clearly shows the effectiveness of our algorithm against false alarms while Table 2 indictes that our algorithm also works effectively during extreme conditions such as breakout. Similar results were obtained for the remaining 8 datasets. We now compare the performance of these algorithms on real datasets. Both the third and fourth datasets are aperiodic and are shown in Fig. 7.

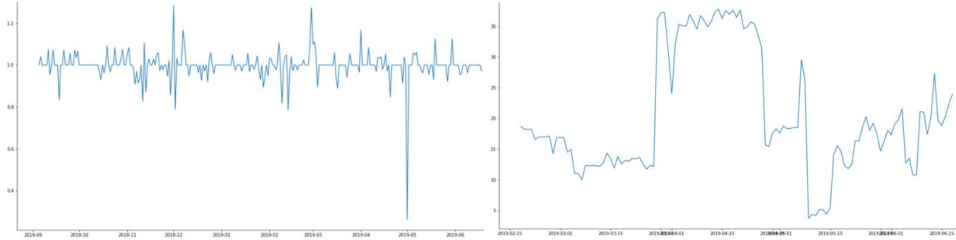

**Fig. 7.** The illustrastion on the left shows the third dataset and the illustration on the right shows the fourth dataset.

**Table 3.** Performance analysis of various algorithms in both online and offline environments by F1 score comparison on the third dataset.

| Algorithm | True Positives | True Negatives | False Positives | False Negatives | F1 Score | Type |
|---|---|---|---|---|---|---|
| TOAD | 13 | 273 | 1 | 0 | 0.96 | OFFLINE |
| Holt Winter | 12 | 274 | 0 | 1 | 0.96 | OFFLINE |
| SARIMA*** | 9 | 272 | 2 | 4 | 0.75 | OFFLINE |
| OC SVM**** | 12 | 272 | 2 | 1 | 0.89 | OFFLINE |
| Prophet | 3 | 272 | 0 | 12 | 0.33 | OFFLINE |
| TOAD | 13 | 271 | 1 | 2 | 0.90 | ONLIN |
| Holt Winter | 12 | 268 | 5 | 2 | 0.77 | ONLIN |
| SARIMA*** | 13 | 267 | 5 | 2 | 0.79 | ONLIN |
| OC SVM**** | 14 | 269 | 3 | 1 | 0.87 | ONLIN |
| Prophet | 3 | 272 | 0 | 12 | 0.33 | ONLIN |

---

\* Here, we have empolyed SARIMA$(1,1,1)(0,0,0)_0$ model.
\*\* We have used ν=0.01 for OC SVM model.
\*\*\* Here, we have empolyed SARIMA$(1,1,1)(0,0,0)_0$ model.
\*\*\*\* We have used ν=0.05 for OC SVM model.

**Table 4.** Performance analysis of various algorithms in both online and offline environments by F1 score comparison on the fourth dataset.

| Algorithm | True Positives | True Negatives | False Positives | False Negatives | F1 Score | Type |
|---|---|---|---|---|---|---|
| TOAD | 5 | 115 | 1 | 0 | 0.91 | OFFLINE |
| Holt Winter | 0 | 116 | 0 | 5 | 0 | OFFLINE |
| SARIMA* | 4 | 114 | 1 | 1 | 0.8 | OFFLINE |
| OC SVM** | 1 | 110 | 5 | 5 | 0.16 | OFFLINE |
| Prophet | 1 | 114 | 2 | 4 | 0.25 | OFFLINE |
| TOAD | 6 | 114 | 0 | 1 | 0.92 | ONLINE |
| Holt Winter | 1 | 105 | 9 | 6 | 0.11 | ONLINE |
| SARIMA* | 5 | 113 | 1 | 1 | 0.83 | ONLINE |
| OC SVM** | 2 | 105 | 9 | 5 | 0.22 | ONLINE |
| Prophet | 1 | 112 | 2 | 6 | 0.20 | ONLINE |

**Table 5.** This Table shows the reduction in time complexity by optimising the number of runs of the algorithm on various datasets.

| Dataset Number | Total number of possible runs | Number of runs before optimisation | Number of runs after optimisation | Total time taken before optimisation | Total time taken after optimisation | Type |
|---|---|---|---|---|---|---|
| Dataset 1 | 265 | 265 | 237 | 35.7 | 31.9 | ONLINE |
| Dataset 2 | 480 | 480 | 261 | 119 s | 41 s | ONLINE |
| Dataset 3 | 267 | 267 | 221 | 34.7 s | 26.6 s | ONLINE |
| Dataset 4 | 101 | 101 | 42 | 7.41 s | 2.72 s | ONLINE |

## 6 Conclusion

This paper proposes a novel algorithm for real time anomaly detection using curve fitting approach and proves its superiority with respect to its existing counterparts. There are many interesting directions for future work. For instance, the problem of automated determination of hyperparameters can be investigated, the effect of public holidays to give context to the anomalies can be studied or criteria for classification based on the severity of anomaly can be considered.

---

* We have employed SARIMA$(1,0,1)(0,0,0)_0$.
** Here, we have taken ν=0.05 for OC SVM model.
*** This algorithm was run on Intel® Core™ i3-7100U CPU @ 2.40GHz × 4 with 8 GB RAM